\documentclass[a4paper,11pt]{article}
\usepackage{amsfonts,amsmath,amssymb,bm,url}
\usepackage{graphicx}
\usepackage{epstopdf}
\usepackage{xcolor}  
\graphicspath{{./CrustcoreB_figs/} }
\usepackage{lipsum}
\usepackage{float}
\usepackage{subfigure}
\usepackage{cleveref}
\usepackage[utf8]{inputenc}
\usepackage{epsfig}

\title{Signatures of strangeness in neutron star merger remnants}

\author{Krishna Prakash Nunna and Sarmistha Banik \thanks{sarmistha.banik@hyderabad.bits-pilani.ac.in} $\>$   \\
Birla Institute of Technology and Science Pilani, Hyderabad Campus,\\
Hyderabad - 500078, India \\
	\and 
	Debarati Chatterjee \\
    Inter-University Centre for Astronomy and Astrophysics, \\
Pune University Campus,\\
Pune - 411007, India
	}

\date{}

\begin{document}

\maketitle
\begin{abstract}
    Neutron star (NS) mergers provide us with information rich in physics using multi-messenger astrophysical observations. One of the probable remnants of such a merger is a differentially rotating hot hypermassive neutron star. The stability of the merger remnant depends crucially on the underlying Equation of State (EoS) and thus provides a method to probe the nature of dense matter in NSs. In this work, we search for possible signatures of strangeness containing matter in the NS interior on the secular stability of the merger remnant. We also use recently proposed methods to make a rough estimate the collapse time of the merger remnant and the threshold mass above which the merger promptly collapses to a black hole.
\end{abstract}

\section{Introduction}
\label{sec:intro}

Neutron stars (NSs) are compact stellar remnants left behind at the end point of evolution of massive stars that end in supernova explosions. Typically having masses of 1-2 solar masses ($M_{solar}$) enclosed within a compact radius of only about 10 km, NSs span a wide range of densities. The interior composition of NSs is still a mystery, as the nature of cold and dense matter beyond saturation density is not accessible to terrestrial experiments, and one must resort to theoretical models for their description. While nuclear experiments provide clues about the nature of matter close to nuclear matter saturation density $n_0$, heavy ion collisions provide information about hot and dense matter at a few times $n_0$. Although such experiments can help to constrain parameters of theoretical models, they must be extrapolated to the regime of low temperatures, higher densities and finite neutron-proton asymmetry to describe NS matter. 
\\

 Strangeness is well established in heavy-ion experiments, where strange particles (hyperons, kaons) have been observed to appear for brief intervals of time. The high densities in the NS core are believed to favour the appearance of strange particles (hyperons, condensates of mesons or even deconfined quarks), which could then exist as stable constituents due to chemical equilibrium via non-leptonic weak interaction processes. The appearance of such additional degrees of freedom should result in reduction of the pressure and consequently a softer Equation of State (EoS) or pressure-density relationship.
 \\
 
 The microscopic EoS of dense matter is one of the key ingredients that govern global astrophysical NS observables, such as its mass, radius or moment of inertia. Thus NS observations can help to constrain its internal structure and composition and hence the EoS of dense matter. For example, solving equations of hydrostatic equilibrium, one can obtain the mass and radius of a NS given its EoS. A softer EoS implies lower pressure at a given density and therefore result in lower NS mass. This would however be incompatible with the recent observation of large NS masses $\sim 2 M_{solar}$ \cite{Demo, Antoniadis}. There have been many suggestions in the recent past to solve this apparent dilemma \cite{Chatterjee2011, Dexheimer, Bednarek, Weissenborn2012a, Weissenborn2012b, LopesMenezes,  Oertel2015, Maslov, CharBanik, Yamamoto}, which revealed the unforeseen role played by interactions among strange particles.
\\

 Apart from NS masses, there are several other observational signatures of strange matter in NSs. The recently launched NICER (Neutron star Interior Composition ExploreR) mission aims to measure radii upto 5$\%$ precision in the near future, and has already started providing interesting constraints on the dense matter EoS \cite{Raaijmakers}. Estimates of NS radii have also been obtained in the recent past from the double pulsar J0737-3039 system \cite{Raithel}, which is well constrained from the measurement of its post-Keplerian parameters. 
 \\
 
 One of the most promising tools that has emerged in the recent past is that of oscillation modes in NSs that emit Gravitational Waves (GWs). Unlike electromagnetic signals that are related to surface phenomena, GWs can directly probe the NS interior composition. Several studies have shown \cite{Chatterjee2006, Chatterjee2007, Chatterjee2008, Chatterjee2009} that unstable modes such as $r$-modes and $w$-modes contain signatures of strange matter in the NS core that can be extracted from the GW signal when detected. Recently, the detection of GWs from the NS binary merger GW170817 has opened up a new window to the universe. Tidal deformations of the NSs in the binary have been used to provide constraints on the NS radius, and consequently on the dense matter EoS \cite{Abbott}.
\\

 The outcome of the NS merger in GW170817 is highly debated, given the uncertainties associated with the detection of the post-merger GWs. 
 Post-merger searches by the LIGO-VIRGO collaboration did not find evidence for GW from the remnant \cite{Abbott2017, Abbott2019a, Abbott2019b}. One possible outcome is a differentially-rotating hot hypermassive NS \cite{Baiotti}. 
The stability of the conjectured hypermassive merger remnant is extremely interesting as it depends crucially on the dense matter EoS as well as the differential rotation velocity profile. Several works in the literature have explored the equilibrium solutions of differentially rotating NSs \cite{Baumgarte00,Rosinska2017}.  
\\

Recently Bozzola et al. \cite{Bozzola2017} and Weih et al. \cite{Weih} have performed studies of the secular instability in hypermassive NSs and proposed a ``quasi-universal" relation between the maximum mass of the remnant and its scaled angular momentum independent of the EoS. The EoSs considered for these works were polytropes, zero-temperature hadronic EoSs or strange star EoSs. Several recent investigations have also probed the threshold mass beyond which the merger remnant collapses to a black hole and the collapse time. However, their estimates and methodology vary widely \cite{Gill,Koeppel,Lucca,Radice}. Further, many of the assumptions that go into such calculations (e.g. slow rotation, spindown via electromagnetic radiation only, consistent treatment of thermal contribution in the EoS) must be carefully reconsidered.
\\

In this work, we investigate the role of strangeness in the NS core on the stability of the hypermassive NS merger remnant. We consider only the most realistic solutions of differentially rotating stars that belong to the class ``A" \cite{Ansorg}, which always have  a mass-shedding limit. The signature of the presence of strangeness containing matter such hyperons and antikaon condensates, on the secular instability is investigates, as well as the universality of the proposed relations. We consider differential rotation and include thermal effects, which are crucial properties of a NS merger remnant. We also estimate the threshold mass of the merger remnant for prompt collapse to a black hole and the corresponding collapse time.  
\\

The outline of the paper is as follows. In Sec.~ \ref{sec:formalism}, we present the formalism, namely the microscopic description and the different EoSs employed in this investigation. We then discuss the numerical scheme employed to obtain the macroscopic NS structure. 
In Sec.~\ref{sec:results}, we obtain the main results of this study.
We provide the details of the method used to determine the onset of the secular instability, investigation of the universal relations and an estimation of the collapse time. Sec.~\ref{sec:discussions} we discuss the main findings of this work and the limitations and scope for further study. 

\section{Formalism}
\label{sec:formalism}

\subsection{Equations of State}
\label{sec:eos}
One of the main goals of this work is to find signatures of strangeness-containing constituents of the NS interior (such as hyperons or antikaon condensates) on the stability of the NS merger remnant. Further, temperatures of 50-100 MeV can be reached in hypermassive NS merger remnants, implying that thermal effects
on the EoS cannot be neglected. It may have a significant effect on the composition, favoring the production of hyperons or mesons. 
In this work, we consider zero-temperature as well as finite temperature EoSs based on the phenomenological Relativistic Mean Field (RMF) with density-dependent coefficients.  

\subsubsection{Zero temperature EoSs}
\label{sec:coldeos}
We consider the following different compositions for the NS core:\\
(i) pure nucleonic matter (DD2) \\
(ii) matter with $\Lambda$-hyperons ($BHB\Lambda \phi$) \\
(iii) matter with antikaon condensates (DD2-$K^-$).\\
The EoSs used in this work all satisfy the 2$M_{solar}$ constraint \cite{Demo, Antoniadis}.
We briefly recapitulate the EoSs below. 
\\

(i) Pure nucleonic matter consists of an ensemble of nuclei and interacting nucleons in nuclear statistical equilibrium.  Uniform nuclear matter contains neutron, proton and leptons at large densities and are described by an RMF model. Nuclei, on the other hand, are treated as separate particle species, their masses are taken from nuclear structure calculations which are based on the same nuclear Lagrangian density. For a given number density and temperature  the Helmholtz free energy is minimized. Also,  within the model the RMF interactions of the nucleons are coupled to the nuclei via chemical equilibrium \cite{hs1}. The EoS is denoted as DD2. The thermodynamically consistent description with excluded volume corrections takes care of the transition of the non-uniform nuclear matter phase from nuclei to uniform nuclear matter. Though this guarantees a smooth transition between the non-uniform and uniform parts of the EoS, it may have  a 4\% uncertainty in radius calculation, as was emphasized by Fortin et al. \cite{Fortin}. But the determination of maximum mass is not affected by the core-crust matching. 
\\

(ii) The NS core density may exceed a few times $n_0$. At high density major constituents of matter may have strange particles like $\Lambda$ hyperons and/or $K^-$ condensates besides protons, neutrons and leptons. The strange particles are never found to coexist with the nuclei as they appear only at high densities. Therefore, we simply use the non-uniform part of the  DD2 EoS \cite{hs1} following the standard prescription of minimization of free energy, as developed by one of our authors in Banik et. al ~\cite{bhb}. In presence of hyperon-hyperon interaction via $\phi$ mesons, the EoS with $\Lambda$-hyperons is represented by BHB$\Lambda \phi$ \cite{bhb}.
\\

(iii) The antikaons are treated on the same footing as the nucleons following  Ref.~\cite{pons}, here the antikaon-baryon couplings are density independent. The interactions between the constituents of such matter are poorly understood largely due to the lack of experimental data. We consider an optical potential depth of -140 MeV for the $K^-$ nucleon interaction. This EoS is represented by DD2-$K^-$.
\\

\subsubsection{Thermal Effects} 
Newly born protoneutron stars at finite temperature as well as hot merger remnants have been studied elaborately in the literature \cite{Abbott2017}. One may consider either isothermal or isentropic configurations. Isentropic configurations are quantified by the value of entropy per baryon $s$ in units of Boltzmann constant $k_B$. We have seen for a fixed entropy per baryon of $s=2k_B$, a NS can shoot a central temperature of 50-100 MeV \cite{Neelam}. Therefore, it is important to study the role of thermal effects on the stability of the NS. In this study, therefore, we compare two cases:\\
(i) Zero temperature ($s=0$) EoSs as elaborated in Sec.~\ref{sec:coldeos} \\
(ii) The same EoSs with thermal effects included ($s=2k_B$).
\\

\subsection{Numerical scheme}
\label{sec:numerical}
Differential rotation may support hot hypermassive NS merger remnants against collapse. In order to study the stability of the merger remnants, one must obtain equilibrium NS configurations for the EoSs discussed above (Sec.~\ref{sec:eos}). There are already existing numerical schemes that compute equilibrium solutions of uniform and differentially rotating cold NSs, see e.g. the numerical library LORENE \cite{Lorene}. Within this scheme, calculations are performed solving general relativistic equations of hydrostatic equilibrium of rotating, axially symmetric stars. The first attempts towards equilibrium models including thermal effects in uniformly and differentially rotating NSs were introduced by Goussard et al. \cite{Goussard98, Goussard97} for realistic EoSs. It was shown that for finite temperature, the integrability of the equation of stationary motion requires an isentropic (constant entropy) or isothermal (constant temperature) solution. Rapidly (uniformly) rotating hot NS configurations were also computed within this framework in \cite{Marques} for realistic EoSs including hyperons. 
\\

In this study we compute equilibrium solutions of hot (isentropic) differentially rotating NSs within the same numerical scheme. Equilibrium equations are solved with Einstein equations, with the assumptions of stationarity, axisymmetry and circularity (absence of meridional convective currents). An EoS is required to close the system of equations. For finite temperature, the EoS depends on temperature as well as on the particle number densities. The partial differential Einstein equations are solved using a multidomain spectral method \cite{BGSM}. 
\\

In order to investigate the role of differential rotation, we employed the usual KEH \cite{KEH} or $j$-constant rotation law defined by the velocity profile:
\begin{equation}
F(\Omega) = R_0^2 \> (\Omega_c - \Omega) \nonumber 
\end{equation}
where $\Omega_c$ is the central angular frequency and $R_0$ is a free parameter with dimensions of length that determines the degree of differential rotation \cite{Baumgarte00}. We consider the dimensionless parameter $a = R_e/R_0$, where $R_e$ is the equatorial NS radius. Thus the limit of uniform rotation is obtained when $a \to 0$ and increasing $a$ denotes increasing degree of differential rotation. The advantage of the $j$-constant law is that it approximately reproduces the rotation-profile obtained in 2D-simulations, and is a “simple” law, with the minimum number of free parameters.
Although the $j$-constant rotation law is the most widely used, alternative rotation laws have been discussed in the literature \cite{ Bozzola2017, Uryu} and should be investigated. However, such a task is beyond the scope of this paper, and we leave it for a future study. It should be noted that the previously obtained stationary state equilibrium solarutions for hot differentially rotating (proto-)NSs were also investigated by Goussard et al. \cite{Goussard97} using the $j$-constant rotation law. 

\section{Results}
\label{sec:results}

\begin{figure}[t]
    \centering
    \includegraphics[height=10 cm,width=15 cm]{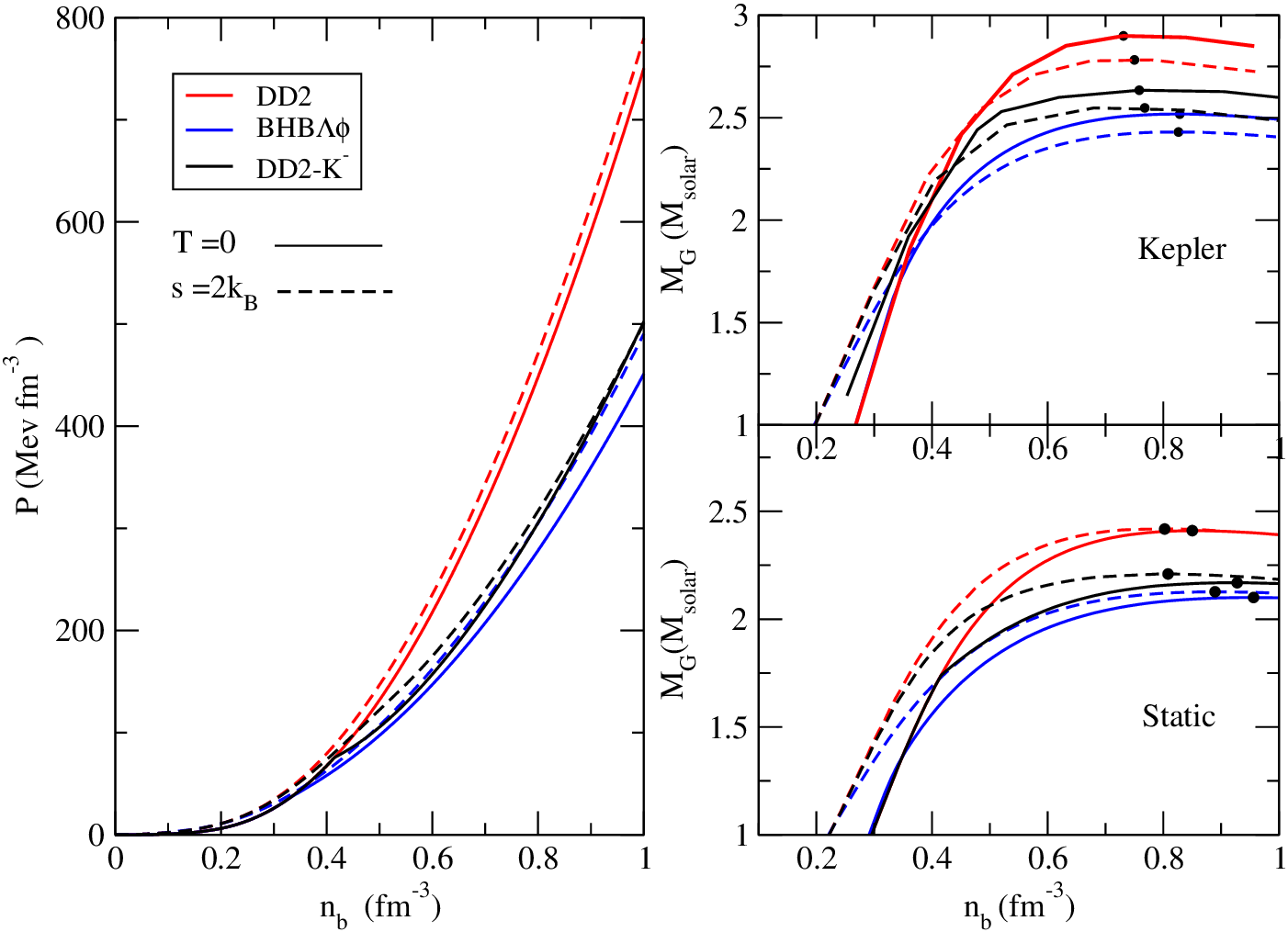}
 \caption{EoSs (left panel) and corresponding gravitational mass -baryon density sequences (static \& uniformly rotating Kepler) are plotted in the right panel. solid lines are used for cold star (T=0), while dashed lines are used for finite entropy per baryon $s=2k_B$. See text for more details. }
    \label{fig:eos_static}
\end{figure}

The EoSs discussed in Sec.~\ref{sec:eos} and the corresponding mass-density relations for the non-rotating  as well as the mass-shedding cases are shown in Fig.~\ref{fig:eos_static}. In the left panel pressure $P$ (in $MeV fm^{-3}$) is plotted against the baryon number density $n_b$ (in $fm^{-3}$) for (i) pure nucleonic matter,  (ii) hyperons 
and (iii) antikaon condensates denoted by DD2, BHB$\Lambda \phi$ and DD2-$K^-$ respectively.
The DD2 EoS is the stiffest of the three, which softens with the advent of extra degrees of freedom in the form of strange particles, $K^-$ condensates and $\Lambda$-hyperons. Again, the EoSs are stiffer for stars with finite entropy per baryon compared to the cold ones for all the three cases, the difference arising from the thermal contribution to the pressure. We use solarid lines for T=0 and dashed ones for stars at $s=2k_B$.
\\

Solving the TOV (Tolman Oppenheimer Volkov) equations of relativistic hydrostatic equilibrium, we obtain the macroscopic structure properties (mass and radius) of the NS. The solutions for the static star corresponding to the different EoSs are plotted in the lower right panel of Fig.~\ref{fig:eos_static}.
As expected, strange EoS yields a lower maximum mass star compared to that of DD2 EoS. A stiffer EoS can support larger mass. However, all the sets of EoS yield maximum mass above the observational 2$M_{solar}$ limit \cite{Demo,Antoniadis}. 
\\

Let us follow  a NS with similar compositions.  Between a cold ($T=0$) and hot ($s=2k_B$) EoS, the latter being stiffer  can support a static star of larger mass, compared to its cold counterpart.
However, this trend is reversed for the mass-shedding sequences. We plot the mass of sequences of uniformly rotating stars at the Keplerian limit in the upper right panel of Fig.~\ref{fig:eos_static}.  The maximum masses for the Keplerian sequences  increase by $\sim$20-23\% more than their static ones for the cold stars, whereas for the stars with $s=2k_B$ the differences are $\sim$12-15\%. The values of maximum mass and other relevant properties are given in Table 1. The thermal pressure contribution can sustain a heavier NS. But it is evident from the table that the Keplerian frequency for a hot NS is not as high as that of the cold star. Hence the difference in maximum mass is lower for hotter stars.


\subsection{Onset of secular instability}
\label{sec:TPcriterion}
The onset of the secular instability is determined by the ``Turning point" (TP) criterion i.e. the maximum of the gravitational masses as a function of central density \cite{Friedman88}. The TP criterion for secular stability in hot rigidly rotating stars was obtained in \cite{Goussard97, Marques} for isentropic (or isothermal) solutions. Considering sequences of differentially rotating equilibrium models using the $j$-constant law, it was shown that a stability criterion for differentially rotating neutron stars exists similar to the one of their uniformly rotating counterparts \cite{Bozzola2017,Weih}. The onset of dynamical instability for differentially rotating stars is marked by the neutral-stability line (where the eigenfrequency of the fundamental mode of oscillation vanishes). The neutral-stability and TP curves coincide for nonrotating neutron stars, but their difference grows with increasing angular momentum. Along a sequence of constant angular momentum, dynamical instability sets in for central rest-mass densities slightly lower than that of secular instability at the TP. 
\\

In order to investigate the effect of the NS core composition on the stability of the NS merger remnant, one must construct relativistic equilibrium sequences and calculate the extra mass supported by the rotating star compared to the static star for the EoSs considered. However it has already been shown that sequences at constant rotation frequency do not allow one to distinguish between stable and unstable solutions \cite{Marques}, but rather sequences of constant angular momentum must be compared. We therefore generate equilibrium sequences at constant angular momentum for given degree of differential rotation $a$.

\begin{figure}[t]
    \centering
    \includegraphics[height=10 cm,width=15 cm]{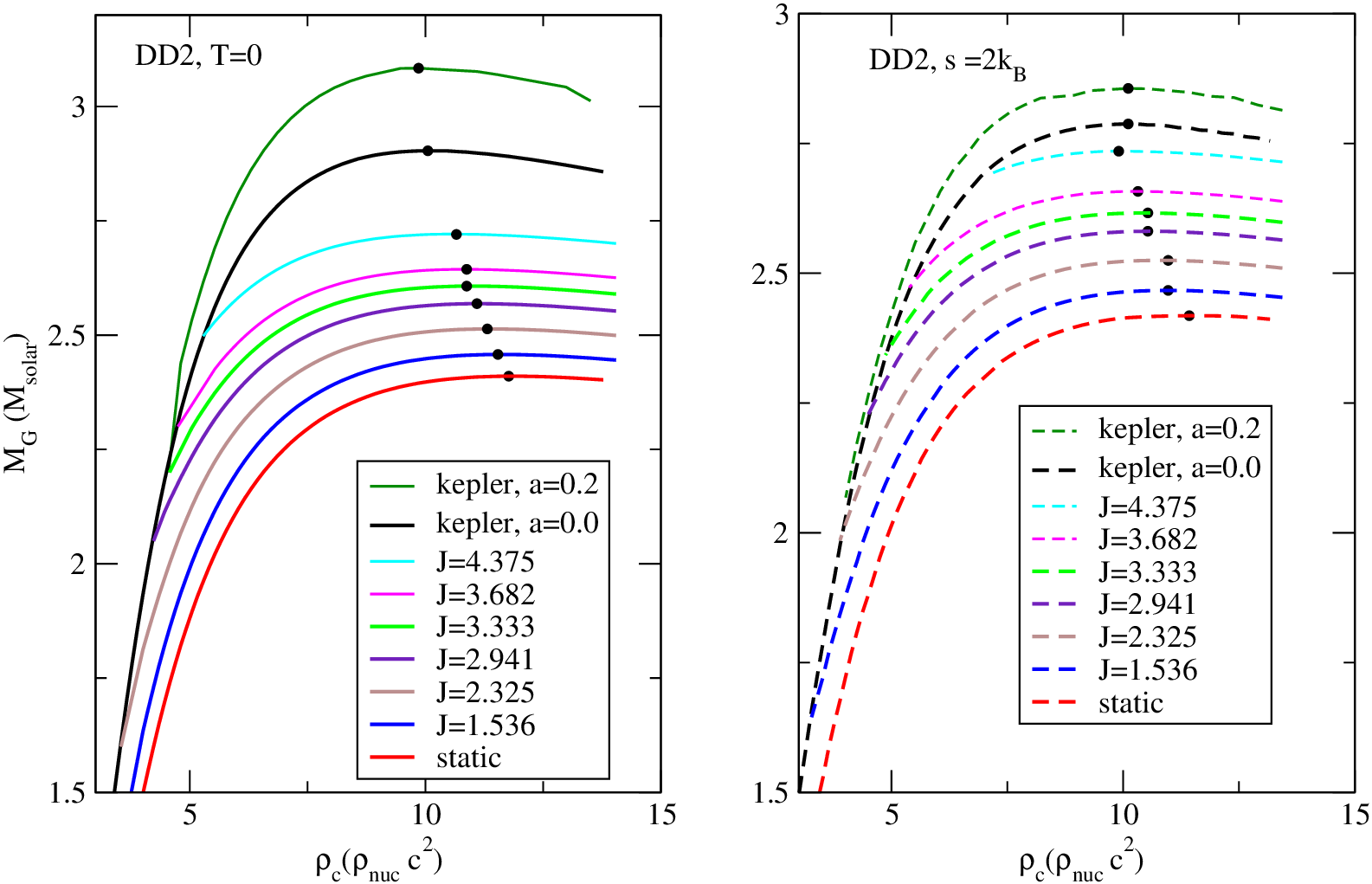}
 \caption{Equilibrium sequences for nucleonic EoS for nonrotating (red ``static" curve) and mass-shedding or ``kepler" limits of uniformly rotating NSs (black curve) and differentially rotating NSs at degree of differential rotation $a=0.2$ (dark-green curve). Also plotted (in colour) are constant angular momentum sequences (labeled by their ``J" values) for $a=0.2$.  Left panel is for $T=0$, right panel is for entropy per baryon $s=2k_B$. The black dots denote the turning points.}
    \label{fig:jconst_rotdiff_np}
\end{figure}

In Fig.~\ref{fig:jconst_rotdiff_np}, we display gravitational mass $M_G$ (in solar masses) as a function of central energy density $\rho_c$ (in units of $\rho_{nuc} c^2$, where $\rho_{nuc}=1.66 \times 10^{17} kg/m^3$) for the nucleonic DD2 EoS. The nonrotating limit are denoted by the red ``static" curves while mass-shedding limit by black ``kepler" curves for uniformly rotating NSs. Also plotted (in colour) are constant angular momentum sequences (labeled by their  angular momentum values ``J") for a given degree of differential rotation ($a=0.2$). In order to study the thermal effects, different values of entropy are considered, $T=0$ in the left panel and $s=2k_B$ in the right panel.
\\

\begin{figure} [t]
    \centering
    \includegraphics[height=10 cm,width=15 cm]{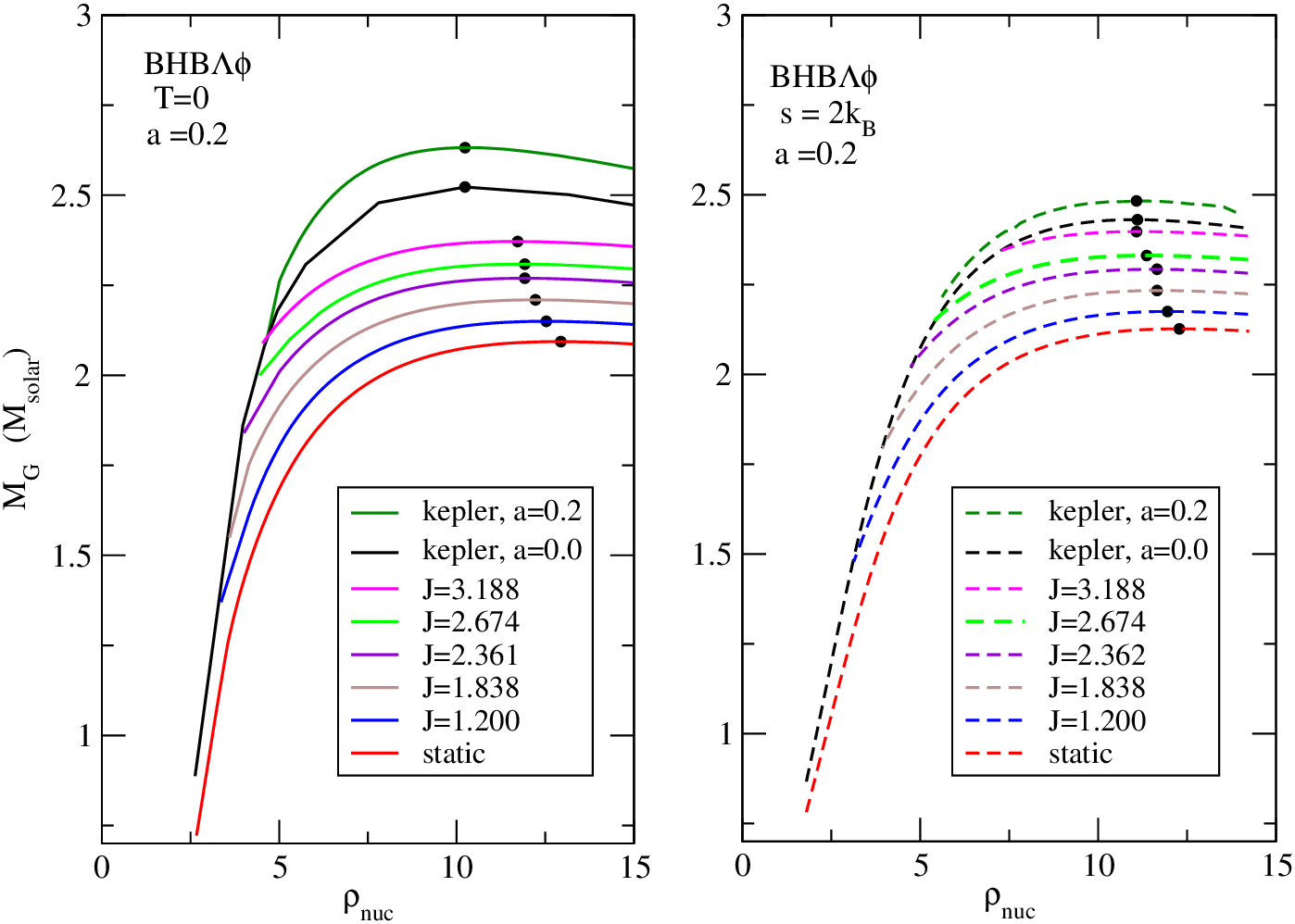}
    \caption{Same as Fig~\ref{fig:jconst_rotdiff_np}, but for BHB$\Lambda\phi$ EoS}
    \label{fig:jconst_rotdiff_nplf}
\end{figure}

In Fig~\ref{fig:jconst_rotdiff_nplf}, we show the static and mass-shedding limit for uniformly rotating equilibrium sequences for $BHB\Lambda \phi$ EoS. Constant angular momentum sequences for differential rotation parameter $a=0.2$ are also included in between the static and Kepler sequences. The left panel displays the zero temperature case while the right panel includes thermal effects ($s=2k_B$). As before, the different parameters are summarized in Table~\ref{tab:tp_nps0_nplf}. 
\\

In Tables~\ref{tab:tp_nps0}, \ref{tab:tp_nps0_nplf} and \ref{tab:tp_nps0_npk}, we study the TP criterion, considering DD2, $BHB\Lambda\phi$ or DD2-$K^-$ EoSs for $s=0$ and $s=2k_B$ respectively. The columns represent respectively the angular momentum $J$ (in GM$^2_{solar}/c$), central energy density $\rho_c$ (in $\rho_{nuc} c^2$), gravitational mass $M_G$ (in M$_{solar}$), central frequency $f_c$ in Hz, ratio of polar and  equatorial radii $r_p/r_e$, ratio of central and equatorial angular frequencies $\Omega_c/\Omega_e$, circumferential radius $R_{circ}$ (in km) and the ratio of kinetic to gravitational energy $T/W$. 
\\

\begin{table*}[htbp]
\begin{tabular}{|c|c|c|c|c|c|c|} 
\hline
\multicolumn{7}{|c|}{ [DD2, $T=0$ ($s=2 k_B$) ]} \\
\hline
 $J$   & $\rho_c$    &  $M_G$   & $f_c$        &   $r_p/r_e$       &     $R_{circ}$   &       T/W  \\
\hline
7.36(5.39)&10.25(10.11)&3.08(2.86)&1820.78(1532.18)&0.48(0.60)&16.51(16.12)&0.167(0.119)\\
4.38&10.46(9.91) &2.72(2.74)&1510.12(1387.25)&0.74(0.70)&13.39(14.97)&0.095(0.093)\\
3.68&10.88(10.32) &2.64(2.66)&1367.54(1272.82)&0.79(0.76)&12.29(14.30)&0.075(0.074)\\
3.33&10.88(10.53)& 2.61(2.62)&1278.70(1191.50)&0.82(0.80)&12.85(13.99)&0.065(0.063)\\
2.94&11.09(10.53) &2.57(2.58)&1176.64(1101.74)&0.85(0.82)&12.64(13.79)&0.054(0.053)\\
2.33&11.32(10.75)& 2.51(2.53)&983.05(925.88)& 0.89(0.88)&12.37(13.42)&0.036(0.036)\\
1.54&11.77(10.97)& 2.46(2.47)&693.18(652.35)&0.95(0.94)&12.0(13.07)&0.017(0.017)\\
\hline
5.91(4.82)&10.05(10.11) &2.90(2.79)&1537.86(1354.41)&0.68(0.58)&13.74(16.92)&0.13(0.10)\\
0(0)&11.77(11.43)& 2.41(2.42)&0(0)&1(1)&11.86(12.72)&0(0)\\

\hline
\end{tabular}
\caption{TP criterion for stars rotating differentially ($a$=0.2) for zero temperature DD2 EoS. The values in parenthesis are for  $s=2k_B$. Among them, the topmost row is for stars spinning at kepler frequency. The last two rows are for uniformly rotating star at Keplerian frequency and for static star respectively.   (See text for details.) }
\label{tab:tp_nps0}
\end{table*}

\begin{table*}[htbp]
\begin{tabular}{|c|c|c|c|c|c|c|} 
\hline
\multicolumn{7}{|c|}{ [$BHB\Lambda\phi$, $T=0$ ($s=2 k_B$) ] } \\
\hline
 $J$   & $\rho_c$    &  $M_G$   & $f_c$        &   $r_p/r_e$       &     $R_{circ}$   &       T/W  \\
\hline
5.02(3.82)&10.24(11.08)&2.63(2.48)&1583.14(1297.45)&0.51(0.54)&16.67(17.48)&0.143(0.103)\\
3.19&11.63(11.08)&2.38(2.40)&1395.82(1303.96)&0.73(0.69)&13.29(14.95)&0.087(0.832)\\
2.67&11.92(11.36)&2.31(2.33)&1272.70(1192.95)&0.79(0.76)&12.85(14.21)&0.069(0.066)\\
2.36&11.63(11.66)&2.27(2.29)&1174.38(1111.55)&0.82(0.80)&12.63(13.81)&0.057(0.055)\\
1.84&12.22(11.66)&2.21(2.23)&988.34(927.662)&0.88(0.86)&12.26(13.37)&0.039(0.037)\\
1.20&12.53(11.95)&2.15(2.18)&698.29(656.40)&0.94(0.93)&11.89(12.91)&0.019(0.017)\\
\hline
4.28(3.44)&10.24(11.10)&2.52(2.43)&1413.10(1277.04)&0.56(0.58)&15.93(16.88)&0.122(0.091)\\
0(0)                &12.84(12.26)& 2.10(2.13)&0(0)&1(1)&11.52(12.5)&0(0)\\

\hline
\end{tabular}
\caption{TP criterion for zero temperature  BHB$\Lambda \phi$ EoS. The values in parenthesis are for  $s=2k_B$  (See text for details.) }
\label{tab:tp_nps0_nplf}
\end{table*}

\begin{table*}[htbp]
\begin{tabular}{|c|c|c|c|c|c|c|} 
\hline
\multicolumn{7}{|c|}{[DD2-$K^-$, $T=0$ ($s=2 k_B$) ] } \\
\hline
 $J$   & $\rho_c$    &  $M_G$   & $f_c$        &   $r_p/r_e$       &     $R_{circ}$   &       T/W  \\
\hline
5.54(4.33)&9.65(10.13)&2.74(2.61)&1586.44(1356.46)&0.50(0.54)&16.94(18.15)&
0.149 (0.109)\\
3.02&11.41(9.86)&2.40(2.44)&1307.14(1140.28)&0.77(0.75)&13.09(14.94)&0.075(0.07)\\
2.89&11.41(10.13)&2.39(2.42)&1273.24(1122.21)&0.79(0.77)&13.00(14.76)&0.070(0.066)\\
1.71&11.96(10.41)&2.26(2.30)&882.16(773.57)&0.90(0.90)&12.23(13.75)&0.031(0.029)\\
1.13&11.94(10.98)&2.21(2.25)&620.09(550.58)&0.95(0.95)&11.94(13.31)&0.015(0.014)\\
\hline
4.65(3.84)&10.13(9.60)&2.62(2.55)&1429.12(1210.26)&0.56(0.58)&16.08(17.79)&0.126(0.095)\\
0(0)                &12.53(10.98)& 2.17(2.21)&0(0)&1(1)&11.69(13.07)&0(0)\\

\hline
\end{tabular}
\caption{TP criterion for zero temperature DD2-$K^-$ EoS. The values in parenthesis are for  $s=2k_B$ (See text for details.) }
\label{tab:tp_nps0_npk}
\end{table*}

\subsection{Universal Relations}
\label{sec:unirel}
\begin{figure} [htbp]
    \centering
    \includegraphics[height=8 cm,width=10 cm]{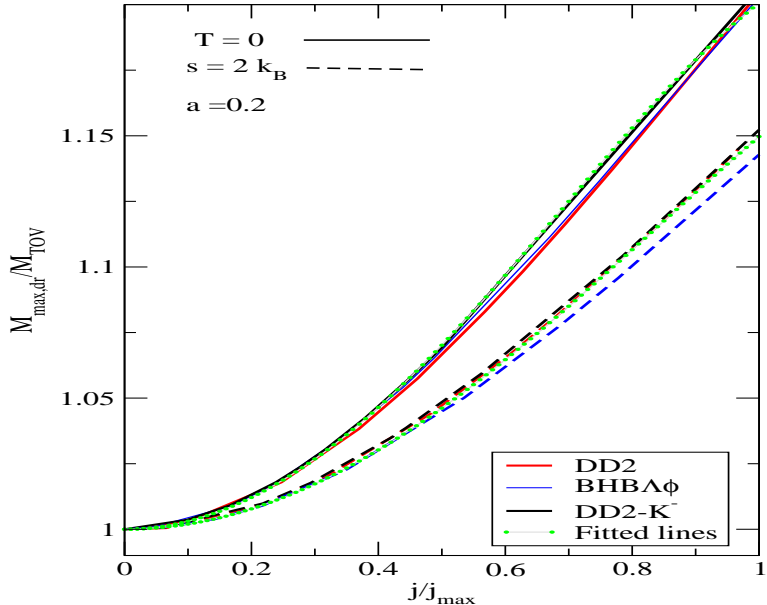}
\caption{Maximum or TP mass normalized to TOV mass of differentially rotating sequences for a given degree of differential rotation ($a=0.2$) as a function of normalized dimensionless angular momentum for different EoSs (see text for details).}
    \label{fig:unirel_a0.2}
\end{figure}

In this work, we investigate whether the presence of strangeness affects the universality of the relations proposed recently \cite{Bozzola2017,Weih}. In Fig~\ref{fig:unirel_a0.2}, the maximum or TP masses of differentially rotating sequences $M_{max,dr}$ for a given degree of differential rotation ($a=0.2$) normalized to the corresponding TOV mass $M_{TOV}$ is plotted as a function of normalized dimensionless angular momentum $j/j_{max}$ for the different EoSs discussed in Sec.~\ref{sec:eos}. Here $j=J/M^2$ and $j_{max}$ is the maximum value of $j$ at the mass-shedding limit for a uniformly rotating NS \cite{Weih}. It is evident from the figure that thermal effects spoil the universality of the relations. This is interesting because the hypermassive NS merger remnant is hot (temperature $\sim$ 80 MeV) and hence thermal effects cannot be ignored. We find that the behaviour of the cold and hot EoSs individually do not vary qualitatively. So we fit the curves for the cold and hot EoSs with a simple polynomial function of the form
\begin{equation} 
    \frac {M_{max, dr}}{M_{TOV}}= 1+ b_1(a)(\frac {j} {j_{max}})^2  + b_2 (a)(\frac {j} {j_{max}})^4
    \label{eq:unirel}
\end{equation}
where the coefficients are found to be $b_1=0.30735(0.1964)$ and $b_2=-0.10671(-0.04671)$ for cold(hot) stars respectively for differential rotation parameter $a$=0.2.

\begin{figure} [htbp]
    \centering
    \includegraphics[height=10cm,width=10 cm]{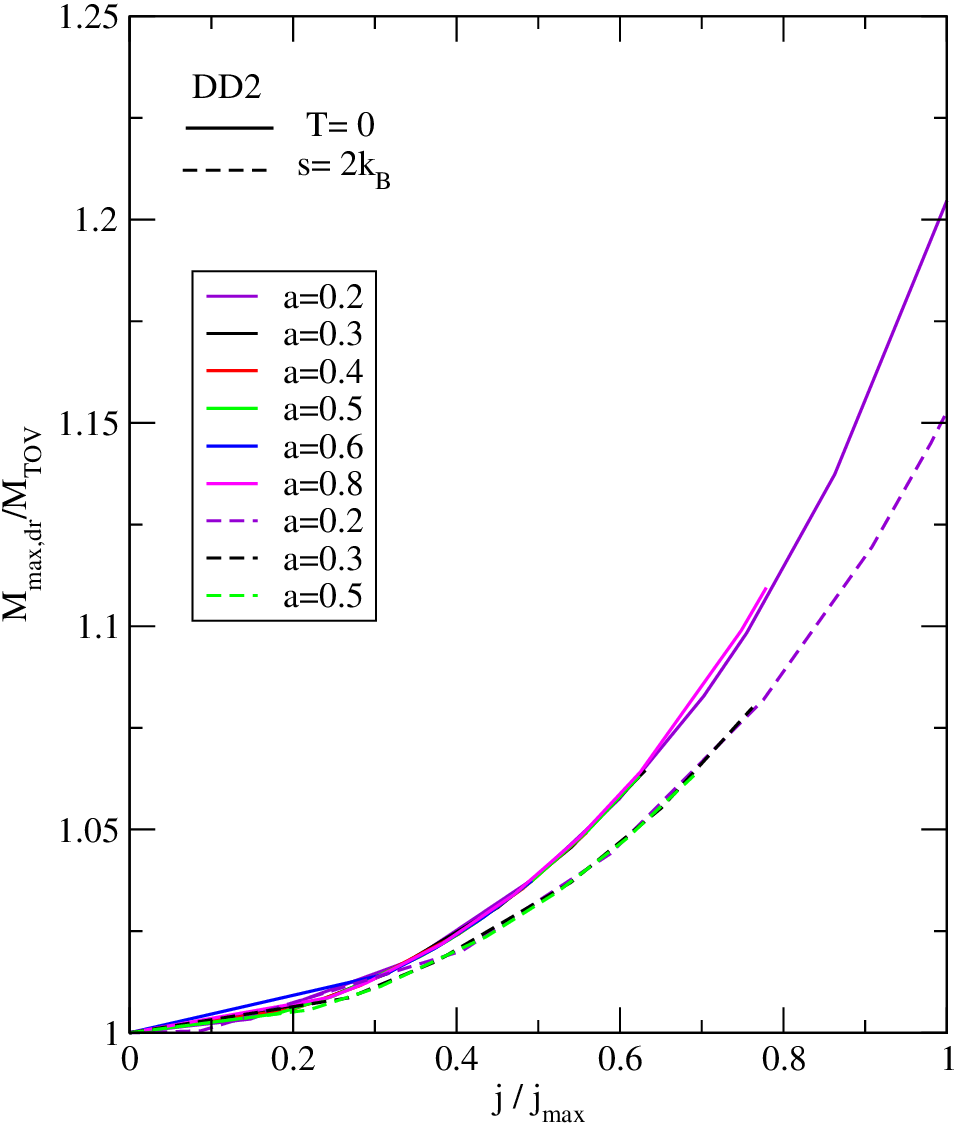}
\caption{Maximum or TP mass normalized to TOV mass of differentially rotating sequences for different degrees of differential rotation $a$ as a function of normalized dimensionless angular momentum for the DD2 EoS (see text for details).}
    \label{fig:unirel_diffa}
\end{figure}

We have already established that irrespective of the EoS, there are two families of curves of hot ($s=2$) and cold ($T=0$) stars for a differentially rotating star with $a$=0.2. We now investigate whether this holds true for other values of differential rotation $a$. In Fig~\ref{fig:unirel_diffa}, the maximum or TP masses normalised to the corresponding TOV mass of differentially rotating sequences for different degrees of differential rotation $a$ are plotted as a function of normalized dimensionless angular momentum for one representative EOS DD2 for the two families $T=0$ and $s=2$. We find that curves in the two families coincide for all the values of $a$ considered ($0 <a < 1$). This also holds true for the other EoSs considered in this study ($BHB\Lambda\phi$ and DD2-$K^-$). Therefore the fit formula proposed in Eq.~(\ref{eq:unirel}) also holds true for other values of $a$ for the EoSs considered in this work. If we take the limit $j/j_{max} = 1$ in this formula, we then obtain the absolute maximum mass of a hot or a cold differentially rotating star,
which gives the values of $M_{max}/M_{TOV}$ = 1.20 (1.15) for the cold (hot) star. The value corresponding to the cold star is lower than the value 1.54 $\pm$ 0.05 obtained by Weih et al. $\cite{Weih}$  and comparable to the value 1.2 of Bozzola et al. $\cite{Bozzola2017}$.
\\

\subsection{Collapse time of the merger remnant}
\label{sec:tcoll}

The value of the total progenitor mass of the NS binary in GW170817 is derived to be 2.74 $M_{\odot}$. The mass of the remnant of the BNS merger has been estimated to be $\sim$ 2.7-2.8 $M_{\odot}$. There are various possible outcomes of a NS merger that have been conjectured \cite{RaviLasky}: \\
(i) a uniformly rotating stable NS (if the progenitor mass $M_P \leq M_{TOV}$). If the EoS is stiff enough, this scenario could be possible \cite{Ai}. \\
(ii) a uniformly rotating supramassive NS (if $M_P > M_{TOV}$). In this case the remnant survives collapse as long as there is enough centrifugal support from rotation. \\
(iii) a hot differentially rotating hypermassive NS (if $M_P$ is greater than the maximum mass supported by uniform rotation). \\

A dynamically unstable hypermassive NS (HMNS) merger remnant may be supported against collapse by the strong differential rotation and thermal pressure. If the remnant is strongly magnetized (protomagnetars with $B \sim 10^{15} G$), differential rotation is damped on the Alfven timescale ($\gtrsim$ 100 ms). Subsequent dissipation of the differential rotation and thermal energy by neutrinos (on the neutrino cooling timescale $\sim$ s) may result in a collapse of the HMNS merger remnant to a black hole on a timescale $\simeq$ 1 s, depending on the total mass, mass ratio and EoS of the NS binary. This may correspond to the 1.74s delay between the merger chirp signal and GRB170817A. The collapse time of the remnant has important implications on multi-messenger astronomy (electromagnetic, GW or neutrino signal). The observation of the blue kilonova and $La$-rich ejecta associated with GW170817 may indicate the formation of HMNSs \cite{Tong}. In the future GW observations from the post-merger remnant with the third generation of interferometers (LIGO India, Kagra, Einstein Telescope) might help to constrain the EoS of merger remnants.
\\

Numerical simulations indicate that the HMNS is formed after the merger with a rapidly rotating highly non-axisymmetric bar-like structure \cite{Gill}. This should result in a time-varying quadrupole moment, with strong emission of GW dominating the spindown. Once the differential rotation is damped, the HMNS may become a supramassive NS configuration, with a spindown dominated by magnetic braking. If the merger remnant is a supramassive NS, the large rotational energy released in the isotropic MHD wind produce a large spin down luminosity $L_{SD} > 10^{42}$ erg s$^{-1}$. But the observed bolometric luminosity is lower than $10^{42}$ erg s$^{-1}$ and no afterglow emission has been seen. Hence the possibility of a supramassive NS remnant may be ruled out. Thus the only possibility to be considered would be that of a hypermassive NS merger remnant.
\\

As we are interested in the stability of the HMNS merger remnant, we would like to make an estimate of the collapse time for the EoSs investigated in this work. However, the collapse time estimates and calculations that exist in the literature vary widely in their formalism and predictions and are far from reaching a consensus \cite{Gill,Koeppel,Lucca,Radice}.
We follow some of the recent suggested methods to obtain estimates of the collapse time and the threshold mass for prompt collapse for the EoSs considered in this work.
\\

 Assuming slow rotation and spindown of a possible SupraMassive NS (SMNS) merger remnant via electromagnetic radiation, \cite{RaviLasky,Lasky} analytically obtained estimates for collapse time using observations of SGRBS (short $\gamma$-ray bursts) by {\it Swift} telescope. However as observational evidence \cite{Gill} now points to the fact that the remnant of the merger GW170817 may be a rapidly differentially rotating HMNS spinning down via gravitational radiation rather than a SMNS, the validity of such relations become questionable.
\\

In the recent work of K\"oppel et al.~\cite{Koeppel}, collapse times were computed using hydrodynamical simulations for five zero-temperature EoSs, adding a ``thermal contribution" via an ideal-fluid EoS \cite{Zanotti}. But this approach of ``hybrid EoS" is known to be non self-consistent. Among the EoSs considered were DD2 and $BHB\Lambda \phi$, which we have also been employed in this investigation for both zero temperature ($s=0$) and finite temperature ($s=2 k_B$). The calculation of threshold mass $M_{th}$ above which the merger remnant promptly collapses to form a black hole was also explored in this work. Extending a previously proposed linear EoS-independent universal relation ~\cite{Bauswein2013,Bauswein2017}
$$M_{th}/M_{TOV}=3.38 \> C_{TOV} + 2.43$$  where the compactness $C_{TOV}=M_{TOV}/R_{TOV}$, they suggested the following non-linear fit formula
$$M_{th} = a - \frac{b}{1-c \> C_{TOV}}$$ where $b=1.01$, $c=1.34$ and $a=\frac{2b}{2-c}$, taking into account the expected black hole limit $M_{th}/M_{TOV} \to 0$ for $C_{TOV} \to 1/2$.
\\

Recently, the results from simulations (see e.g. \cite{Koeppel}) and observations (e.g. \cite{Lasky,RaviLasky}) were combined to derive a radius-independent fit relation between the initial mass of the single NSs $M_{NS}$ and the collapse time $t_{coll}$ \cite{Lucca} :
\begin{equation}
\log(t_{coll}) = e_0 + e_1 \log \left( \frac{M_{NS}}{M_{TOV}} \right)~,
\label{eq:fit_lucca}
\end{equation}
where $e_0=-5.45\pm0.40$ and $e_1=-38.9\pm1.7$.
The robustness of such a relation was tested and proposed as a useful tool to constrain NS EoSs ~\cite{Lucca}. $BHB\Lambda \phi$ ($s=0$) EoS was also one of the EoSs considered.
\\

In order to compare with the results discussed above, we consider the same values of initial data as in \cite{Koeppel} namely initial masses $M_{NS}$=1.53, 1.55 and 1.57 $M_{solar}$ for the DD2 EoS and 1.62,1.63 and 1.65 $M_{solar}$ for $BHB\Lambda\phi$ EoS. Using the formula for $M_{th}$ suggested in \cite{Koeppel}, we calculated the threshold mass for prompt collapse in units of $M_{TOV}$.
The results for $s=0$ as well as $s=2$ case are summarized in Table \ref{tab:tcoll}.
For comparison with previous works, we also provide the maximum  static  mass $M_{TOV}$,  corresponding  radius $R_{TOV}$, compactness $C_{TOV}$ and the free-fall timescale
\begin{equation}
\tau_{TOV} = \frac{\pi}{2} \sqrt \frac{R_{TOV}^3}{2 M_{TOV}}~.
\end{equation}
We then apply Eq.(\ref{eq:fit_lucca}) to calculate collapse times  $t_{coll}$ for comparison with the results of \cite{Lucca}. The estimated values of $t_{coll}$ (taking only the mean values for the fit coefficients $e_0$ and $e_1$) corresponding to the different $M_{NS}$ are provided in the Table~\ref{tab:tcoll}. 
\\

\begin{table}[ht]
 \caption{For each EoS, we provide the maximum static mass, corresponding radius, compactness, free-fall timescale \cite{Koeppel}, threshold mass for prompt collapse \cite{Koeppel} and collapse time \cite{Lucca}. See text for more details. }
    \centering
    \begin{tabular}{|c|c|c|c|c|c|c|c|c|}
\hline
    $EoS$ & $M_{NS}$ & $s$ & $M_{TOV}$ & $R_{TOV}$ & $C_{TOV}$ & $t_{coll}$ & $\tau_{TOV}$ & $M_{th}$ \\
    {} & $(M_{solar})$ &($k_B$)&$(M_{solar})$ & (km) & {} &(ms)& $({\mu s})$ & $(M_{TOV})$ \\
    \hline
    DD2 & 1.62 & 0 & 2.41 & 11.86 & 0.30 & 18.2 & 80.2 & 1.37\\
    &1.63 & &  &  &  & 14.29 & &\\
     &1.65 & &  &  &  & 8.89 & &\\
     \hline
   DD2    & 1.62 & 2 & 2.418 & 12.72 & 0.28 & 20.72 & 88.9 & 1.44\\
       & 1.63 & & & & & 16.31 & &\\
       & 1.65 & & & & & 10.15 & &\\
       \hline
$BHB\Lambda\phi$ & 1.53 & 0 & 2.10 & 11.52 & 0.27 & 0.79 & 82.26 & 1.48\\
               &1.55      &   &      &       &      & 0.48 &       &      \\
               &1.57      &   &      &       &      & 0.29 &    &       \\
               \hline
$BHB\Lambda\phi$ & 1.53 & 2 & 2.127 & 12.5 & 0.25 & 1.31 & 92.36 & 1.54 \\
               &  1.55   &   &      &       &      & 0.79 &       &      \\
               &    1.57  &   &      &       &      & 0.48 &       &       \\

 \hline
\end{tabular}
\label{tab:tcoll}
\end{table}

\section{Discussions}
\label{sec:discussions}

Since the detection of GWs from the NS binary merger event GW170817, the fate of the binary remnant remains a mystery. As no evidence of a remnant has yet been found from post-merger searches by the LIGO-VIRGO collaboration, one may study the different possibilities theoretically. One likely outcome of the merger is a metastable differentially rotating hot hypermassive neutron star. As the stability (dynamical and secular) and time of subsequent collapse of the remnant depend on its rotation profile and its interior composition, it opens the possibility to constrain the dense matter EoS from its stability analysis.
\\

In this work, we explored the onset of secular instability for different EoSs with and without strangeness. Using the Turning Point (TP) criterion, we investigated the maximum mass that may be supported by differential rotation and thermal effects for the different EoSs considered. We found that inclusion of thermal effects reduced the maximum mass of the differentially rotating configurations. This is interesting as the hypermassive remnant is conjectured to be hot, and hence thermal effects cannot be neglected. 
When studying the maximum mass supported by a hypermassive NS remnant, previous works considered cold stars or a very restricted
sets of EoS, e.g. polytropic EoSs or only nucleonic matter. With realistic EoSs including hyperonic and kaonic degrees of freedom we investigated the influence of these new degrees of freedom on the maximum supported mass.
\\

We found that the maximum mass obtained depends both on the EoS and the degree of differential rotation. In order to calculate the highest possible value of the maximum mass, we followed the method for obtaining a ``universal relation" proposed by \cite{Baiotti} for uniform rotation, extended for the case of differential rotation by \cite{Bozzola2017,Weih}. However for the EoSs considered, we found the universal relation to be practically independent of the EoS and the degree of differential rotation. The highest mass obtained in our analysis was $M_{max}$ = 1.2 $M_{TOV}$ for cold NSs and 1.15 $M_{TOV}$ for hot NSs.
\\

We further investigated the effect of strangeness on the collapse time of the merger remnant. We considered the scenario in which the hypermassive NS merger remnant rapidly loses angular momentum due to loss of energy by GW emission and collapses to a black hole before the Alfven timescale, i.e. before the differential rotation is damped by magnetic dissipation. This scenario is currently favoured by the combined multi-messenger astrophysical observations \cite{Gill}. We estimated the collapse time and threshold mass for prompt collapse for the EoSs with and without strangeness, using recently proposed fit formulas \cite{Koeppel,Lucca} obtained using observations of short gamma ray bursts \cite{Lasky} and hydrodynamical simulations \cite{Koeppel}.
\\

Post-merger MM searches may be able to answer the question about the fate of the merger remnant in GW170817, by ruling out some of the proposed scenarios. Future GW events from other NS mergers along with MM observations will provide further information about the stability of NS merger remnants as well as the dense matter EoS. An exciting journey in MM astronomy has only begun. 
\\
 
\section{Acknowledgements}
\label{sec:ack}
DC would like to thank the warm hospitality of BITS Pilani Hyderabad campus and Saha Institute of Nuclear Physics where part of this work was completed. SB is grateful to Micaela Oertel for insightful discussions.
\\

\end{document}